\def\year{2019}\relax
\documentclass[letterpaper]{article} % DO NOT CHANGE THIS
\usepackage{aaai19}  % DO NOT CHANGE THIS
\usepackage{times}  % DO NOT CHANGE THIS
\usepackage{helvet} % DO NOT CHANGE THIS
\usepackage{courier}  % DO NOT CHANGE THIS
\usepackage[hyphens]{url}  % DO NOT CHANGE THIS
\usepackage{graphicx} % DO NOT CHANGE THIS
\def\UrlFont{\rm}  % DO NOT CHANGE THIS
\usepackage{graphicx}  % DO NOT CHANGE THIS
\newcommand*{\ie}{{\em i.e.,~}}
\newcommand*{\eg}{{\em e.g.,~}}

\newcommand*{\ignore}[1]{} 
\newcommand*{\nop}[1]{}

\title{Multilingual Multimodal Digital Deception Detection and Spread\\ across Social Platforms}

\author{
	Maria Glenski, Ellyn Ayton, Josh Mendoza, and Svitlana Volkova\\
	Data Sciences and Analytics Group \\ 
	Pacific Northwest National Laboratory\\ 
	Richland, Washington 99352\\ 
	\textrm{\{maria.glenski,~ellyn.ayton,~joshua.mendoza,~svitlana.volkova\}@pnnl.gov}
}
   
\begin{document}

\maketitle

\section{Motivation}
 
There is a deluge of information flooding feeds, timelines, and forums online but even in the wake of this, users are increasingly relying on these platforms for \textit{news} and information to base not only their opinions but actions. An August 2018 survey from the Pew Research Center found that 68\% of Americans report that they get at least some of their news from social media~\cite{matsa2018pew} while another found that the rate of Americans who \textit{often} get their news online in some way increased from 38\% in 2016 to 43\% in 2017~\cite{bialik2017pew}. Alongside the increased access to information and on-demand news about local and global events alike, there has been a deluge of misleading or deceptive misinformation. The spread of this ``digital disinformation'' within and across networks is of great concern.

The impact of digital disinformation has been seen in several areas from natural disasters and other crisis events~\cite{starbird2014rumors,starbird2017examining,takahashi2015communicating} to politics~\cite{hadgu2013political}, health-related conspiracies~\cite{seymour2015advocacy,jolley2014effects}, and more. At the level of individual experiences, recent Pew Research Center studies have found that the average user is highly concerned about misinformation in their general use: 31\% see inaccuracy as the top concern when consuming news from social media, 64\% of adults believe fake news stories caused a great deal of confusion~\cite{mitchell2016pew}, and 57\% of social media users who consume news from one or more of those platforms expect the news they see to be ``largely inaccurate''~\cite{matsa2018pew}.

\section{Recent Work}
Many studies focused on digital deception, in particular deception in open-source data including social media, have focused on rumor and misinformation detection with a primary focus on the network's role in information diffusion models~\cite{qazvinian2011rumor,kwon2013prominent,wu2015false,kwon2017rumor}. Other studies compare and contrast the behavior of traditional and alternative media~\cite{starbird2017examining}, classify media sources into sub-categories of misinformation~\cite{wang-2017-liar,perez2015experiments,volkova2017separating,baly2018}, or attempt to detect rumor-spreading users~\cite{rath2017retweet}. 

Our main contribution in this work is novel results of multilingual models that go beyond typical applications of rumor or misinformation detection in English social news content to identify fine-grained classes of digital deception across multiple languages (\eg Russian, Spanish, etc.). In addition, we present models for multimodal deception detection from images and text and discuss the limitations of image only and text only models. Finally, we elaborate on the ongoing work on measuring deceptive content (in particular disinformation) spread across social platforms.

\begin{figure*}[t]
	\centering
	\includegraphics[width=0.75\textwidth]{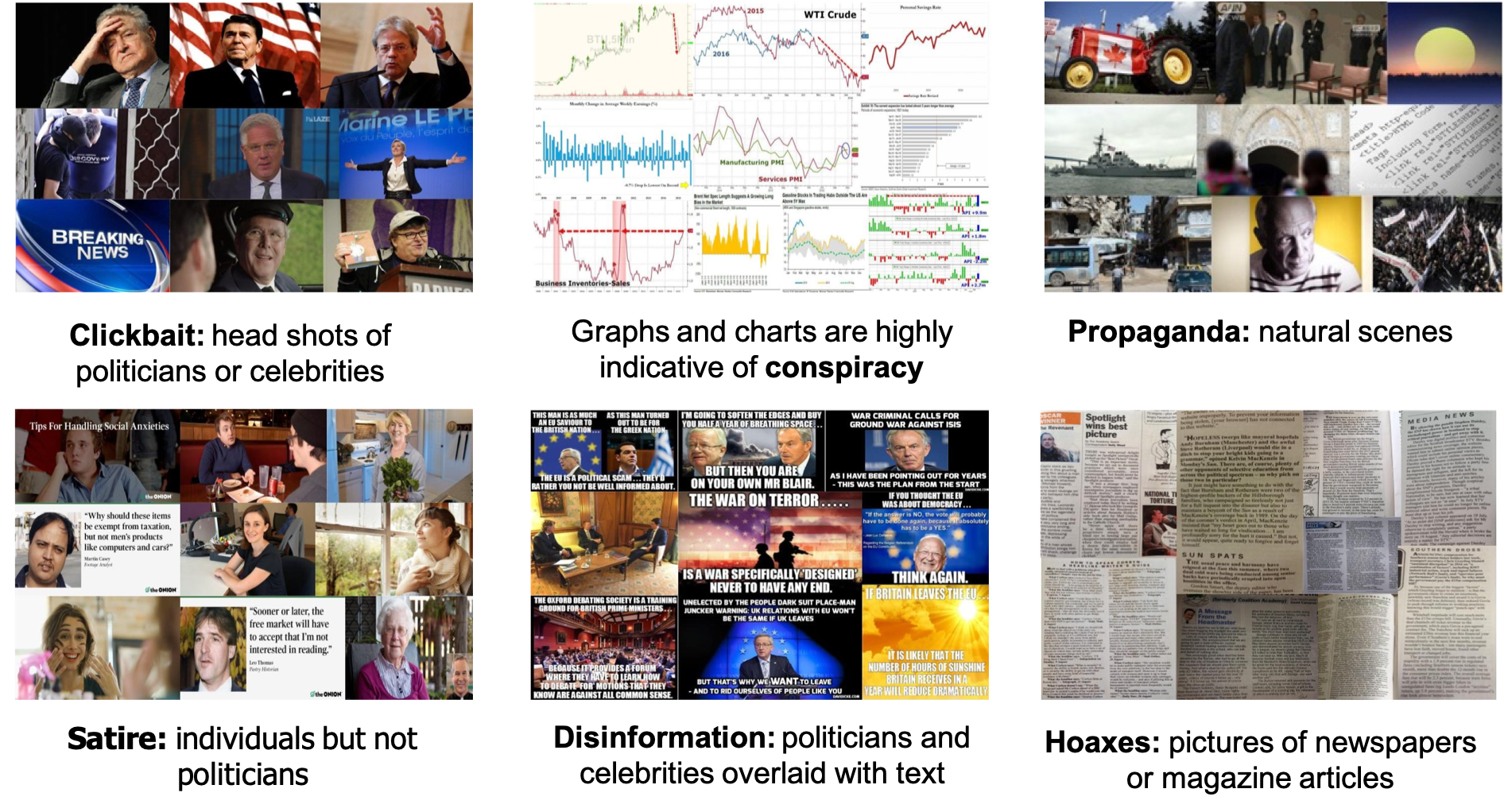}  
	\caption{Key findings from the analysis of multimodal deception detection model behavior from \cite{volkova2019explaining}.}
	\label{fig:multimodal_findings}
\end{figure*}

\paragraph{Identifying linguistic markers of digital deception and their contribution to detection models in English}

First, we examined linguistic cues of digital deception across a spectrum (disinformation, propaganda, conspiracy, hoaxes, and satire) to develop predictive models with which we classified 130 thousand news posts on Twitter as suspicious or verified, and predicted four sub-classes of suspicious news – satire, hoaxes, clickbait and propaganda~\cite{volkova2017separating}. 
Similar to~\citeauthor{vosoughi2018spread}~\citeyear{vosoughi2018spread} and~\citeauthor{grinberg2019fake}~\citeyear{grinberg2019fake}, we relied on source-based annotations rather than content-level annotations to capture the intent behind spreading deceptive content.

Through this study, we identified several key differences between news tweets of different sub-classes when spread online. For example, we saw that credible news tweets (\ie those that did not spread digital deception) contain significantly less
bias markers, hedges and subjective terms and less
harm/care, loyalty/betrayal and authority moral
cues compared to news tweets spreading varied types of digital deception.  

The best performing models built to detect these fine-grained classes achieved accuracy of 0.95 on the binary predictive task and F1 of 0.91 on the multi-class. This work was built upon in a subsequent studies~\cite{rashkin2017truth,volkova2018misleading} that examined linguistic and lexical features of various types of deceptive news articles and disinformation statements to show that these features contribute to the understanding of the differences between reliable news sources and those that spread disinformation which can also be leveraged for effective detection models.

\paragraph{Understanding patterns of engagement and information spread from deceptive news sources} In a large-scale study of news in social
media, we analyzed 11 million posts and investigated the propagation behavior of users that directly interact with news accounts
identified as spreading credible information versus digital deception~\cite{glenski2018propagation}. Unlike
previous work which primarily looks at specific rumors or events,
we focused on news sources in an effort to bridge the gap in the understanding of how users react to {\it news sources} of varying credibility and how their various initial responses contribute the the spread of digital deception.   

Our analysis identified several key differences in propagation behavior from credible versus suspicious news sources such as high inequity in the diffusion rate based on the source of deception, with a small group of highly active users responsible for the majority of disinformation spread overall
and within various demographics. Our demographics-based analysis found that users with lower annual income and education share more from disinformation sources compared to their counterparts.  
In a subsequent study, we further identified significant differences in the patterns of engagements from automated ``bot'' accounts versus humans~\cite{glenski2018humans} and how user-reactions to deceptive and credible news sources remain consistent as well as differ across multiple platforms~\cite{glenski2018identifying}.

\paragraph{Explaining multimodal digital deception detection}
Another recent area of interest has been the incorporation of image-based features to more accurately identify multimodal digital deception. 

In this recent study~\cite{volkova2019explaining}, we presented multi-modal deceptive news classification models and an in-depth quantitative and qualitative analysis of their behavior when classifying various classes of digital deception that incorporates imagery alongside text. Key findings of the analysis of model behavior are highlighted in Figure~\ref{fig:multimodal_findings}. When we compared the performance of models that rely on text or image features alone along with models trained on text and images jointly, we found that the latter models outperform the individually trained model with F1 scores as high as 0.74 for binary classification of deceptive (propaganda or disinformation) versus credible. Our quantitative analysis reveals that when considering only one aspect of the content, the text only models outperform those that just leverage image features (by 3-13\% absolute in F-measures). Finally, we also presented a novel interactive tool ErrFILTER\footnote{ErrFilter is available at \url{https://github.com/pnnl/errfilter}} that allows users to explain model prediction by characterizing text and image traits of suspicious news content and analyzing patterns of errors made by the various models, which can in turn be used to inform the design of future digital deception detection models.

\begin{figure*}[ht!]
	\centering
	\includegraphics[width=0.8\textwidth]{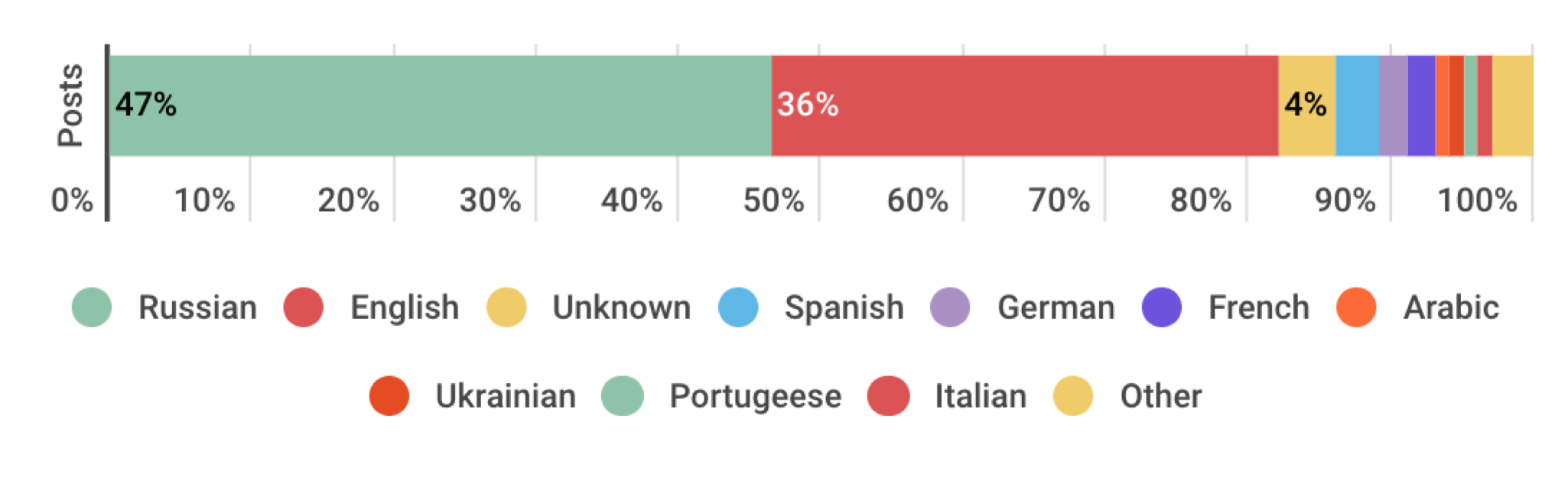}
	\vspace{-\baselineskip}
	\caption{Distribution of languages represented in the large multilingual dataset of 7M Twitter posts from 2016.}
	\label{multilingual_data}
\end{figure*}

\section{Multilingual Deception Detection}
The main contribution of this extended abstract focuses on the results of our predictive models that identify fine-grained classifications of digital deception and related deceptive content from multilingual social media postings. Similarly to the recent work described above that focused on multi-platform and multi-modal deception detection, we focus on multi-dimensional computational approaches to identifying multilingual digital disinformation and misinformation being spread in social media.  

\paragraph{Detection Task} In this work, we concentrate on deception detection in multilingual social media postings structured as two multi-class classification tasks :
\begin{enumerate}
	\item 4-way Classification
	
	\indent Given a social media posting, classify the text as Propaganda, Conspiracy, Hoax, or Clickbait.
	\item 5-way Classification
	
	\indent Given a social media posting, classify the text as Disinformation, Propaganda, Conspiracy, Hoax, or Clickbait.
	
\end{enumerate}

\paragraph{Multilingual Twitter Data} The dataset used in this work comprises 7M posts in a variety of languages from English to Russian, German, and Spanish. We present the distribution of languages represented in this dataset in Figure~\ref{multilingual_data}. Here, we see that the majority (47\%) of the data collected has text written in Russian, followed by English posts (36\%) with the remainder composed of Spanish, German, French, Arabic, Ukranian, Portugeese, Italian, and other languages.

\paragraph{Multilingual Deception Detection} Building off of previous, related tasks~\cite{volkova2017separating}, we use a similar neural architecture for this task. This architecture, composed of a network and/or linguistic cues sub-network (left) and a text representation (as word or characters) sub-network (right).

\begin{figure}[t]
	\centering
	\includegraphics[width=0.45\textwidth]{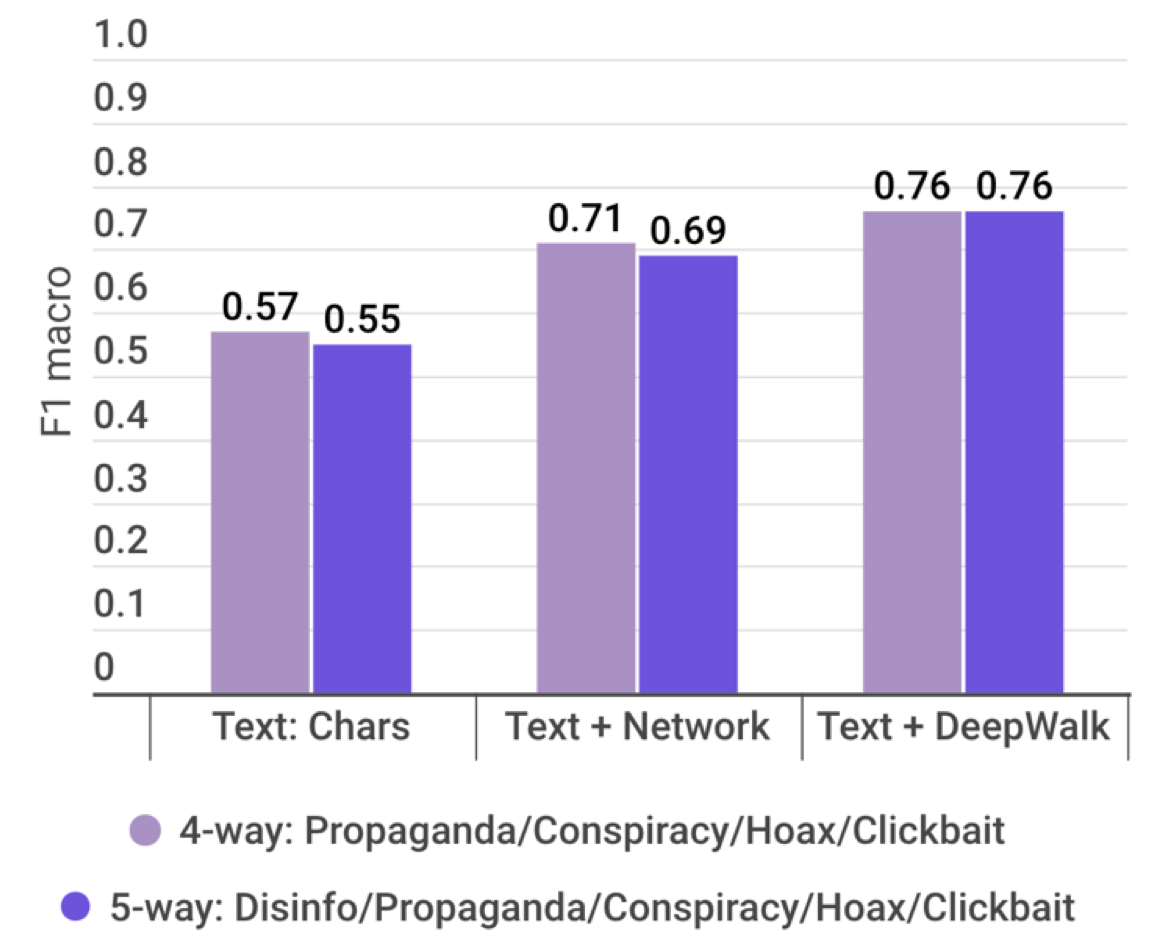}  
	\caption{Macro F1 scores for 4-way and 5-way classification tasks in a multilingual setting.}
	\label{fig:multilingual_f1macro}
\end{figure}

As typical with other multi-classification tasks, we report the macro F1 scores for each of our models on the 4-way and 5-way classification tasks.
We highlight one of our key findings in Figure~\ref{fig:multilingual_f1macro} which illustrates the model performance in terms of macro F1 scores on the multilingual dataset. Unlike for English (explored in \cite{volkova2017separating,rashkin2017truth}), text representations in \textit{characters} in combination with DeepWalk representation network features achieve the best performance of 0.76 for both the 4-way and 5-way classification tasks.

\ignore{
	\begin{figure}[ht!]
		\centering
		\includegraphics[width=0.45\textwidth]{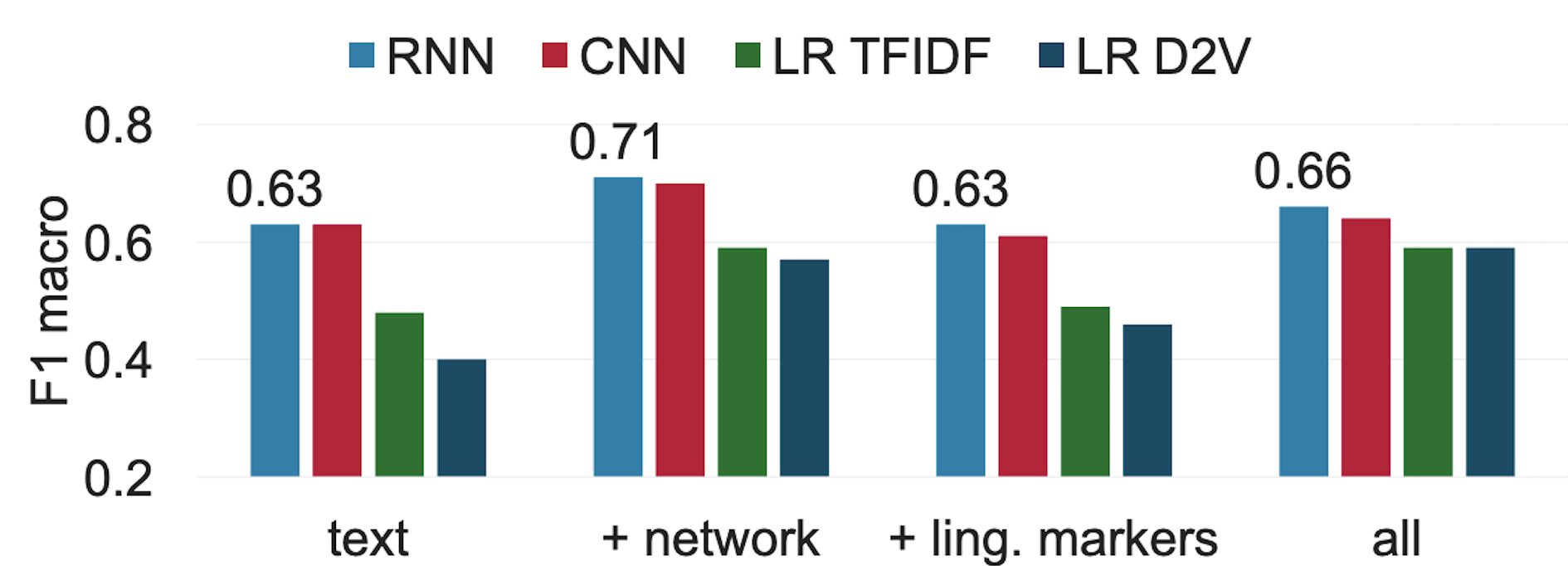}
		\vspace{-\baselineskip}
		\caption{Macro F1 scores for 4-way  classification tasks in a multilingual setting that incorporates Twitter posts in Russian, Spanish, German, French, Arabic, Ukrainian, Portuguese, and Italian as well as English content}
		\label{acl_separating}
	\end{figure} 
}

\section{Ongoing Work:  Measuring Cross-Platform Spread of Disinformation} 

Intuitively, the next steps need to focus on measuring \textit{cross-platform} spread of disinformation. Social media platforms on which digital disinformation is released and promoted do not exist in a vacuum, and, thus, the connections between platforms and communities have a potentially significant impact on the spread and virality of digital disinformation. Although we have previously examined the concurrent spread of information from news sources on multiple platforms, our continued efforts in this area focus on incorporating the cross-platform links between users who actively spread or engage with disinformation and the cross-platform URLs associated with disinformation narratives.

Under DARPA SocialSim program\footnote{https://www.darpa.mil/program/computational-simulation-of-online-social-behavior}, we developed a unified framework for measuring information spread and evolution within and across social platforms, that was presented at the ICWSM 2019 tutorial\footnote{https://sites.google.com/alumni.nd.edu/icwsm19t3/}.Our framework will allow us to measure disinformation spread within and across social platforms e.g., Twitter, Youtube and Telegram focusing on specific social phenomena -- information cascades, recurrence, persistent groups and coordinated effort. More specifically, we will focus on several user cases of known disinformation: the White Helmets\footnote{\url{https://www.washingtonpost.com/world/russian-disinformation-campaign-targets-syrias-beleaguered-rescue-workers/2018/12/18/113b03c4-02a9-11e9-8186-4ec26a485713_story.html}}, Syrian airstrikes\footnote{\url{https://www.foxnews.com/world/russian-trolls-ramp-up-disinformation-campaign-after-syria-airstrikes-pentagon}} and NATO exercises\footnote{\url{https://dod.defense.gov/News/Article/Article/1649146/nato-moves-to-combat-russian-hybrid-warfare/}}.
Additionally, we will be measuring disinformation spread and the effect of censorship during internet outages during crisis events in Venezuela\footnote{\url{https://www.npr.org/2019/01/26/688868687/amid-chaos-venezuelans-struggle-to-find-the-truth-online}} and Sri Lanka\footnote{\url{https://www.washingtonpost.com/technology/2019/04/22/sri-lankas-social-media-shutdown-illustrates-global-discontent-with-silicon-valley/?utm_term=.6166a484a0c7}}.

 \section*{Biographical Notes}
 
 \noindent \textbf{Maria Glenski}  received the Ph.D. degree from the University of Notre Dame, Notre Dame, IN, USA, in 2019. She is currently a Data Scientist in the Data Sciences and Analytics group, National Security Directorate, Pacific Northwest National Laboratory. Her research in social media analysis,  rating systems, and social news consumption has been published in the ACM Conference on Hypertext and Social Media, ACM Transactions on Intelligent Systems and Technology, and the ACM Conference on Computer-Supported Cooperative Work and Social Computing.

 \medskip
\noindent  \textbf{Ellyn Ayton}  
 Ellyn Ayton is a Data Scientist at Pacific Northwest National Lab. Her work focuses on machine learning and natural language processing, with applications to deceptive news detection and social media analytics. Ellyn received her MS in Computer Science from Western Washington University in 2018.
 
 \medskip
\noindent \textbf{Josh Mendoza}  
 Joshua Mendoza is currently a Data Scientist in the Data Sciences and Analytics group, National Security Directorate, Pacific Northwest National Laboratory. His work focuses on deep learning and engineering, with applications in natural language processing and computer vision. Joshua received his bachelor’s degree with honors in 2015 from the University of Washington.
 
 \medskip
\noindent \textbf{Svitlana Volkova} received the Ph.D. degree from Johns Hopkins University, Baltimore, MD, USA, in 2015. She is currently a senior scientist at the Data Sciences and Analytics group, National Security Directorate, Pacific Northwest National Laboratory. Her research focuses on advancing machine learning, deep learning and natural language processing techniques to build novel predictive and forecasting social media analytics. Her models advance understanding, analysis, and effective reasoning about extreme volumes of dynamic, multilingual, and diverse real-world social media data. She was awarded the Google Anita Borg Memorial Scholarship in 2010 and the Fulbright Scholarship in 2008. Dr. Volkova is a Vice Chair of the ACM Future of Computing Academy.
 
\bibliographystyle{aaai}
\bibliography{bibfile,ref}

\end{document}